\documentclass[fleqn]{article}
\usepackage{pdfpages}
\usepackage{graphicx}
\graphicspath{{}}

\usepackage[utf8]{inputenc}
\usepackage[english]{babel}
\usepackage[document]{ragged2e}

\usepackage[fleqn]{amsmath}

\title{Green Function Theory of Strongly Correlated Electron Systems}

\author{Tao Sun %
\thanks{Contact: \texttt{tao@taosphysics.net}}}

\begin{document}

\maketitle

\begin{abstract}

A novel effective Hamiltonian in the subspace of singly occupied states is obtained by applying the Gutzwiller projection approach to a generalized Hubbard model with the interactions between two nearest-neighbor sites.
This model provides a more complete description of the physics of strongly correlated electron systems.
The system is not necessarily in a ferromagnetic state as temperature $T\rightarrow 0$ at any doping level $\delta\geq 0$.
The system, however, must be in an antiferromagnetic state at the origin of the doping-temperature ($\delta$-$T$) plane ($T\rightarrow 0$, $\delta=0$).
Moreover, the model exhibits superconductivity in a doped region at sufficiently low temperatures.
We summarize the studies and provide a phase diagram of the antiferromagnetism and the superconductivity of the model in the $\delta$-$T$ plane here.
Details will be presented in subsequent papers.

\end{abstract}

\section{Introduction}

Since the discovery of high-temperature superconductivity in 1986 \cite{bednorz}, a tremendous number of studies have been devoted to the understanding of its underlying mechanism.
It is generally believed that some essential physics of this strongly correlated electron system can be described by the Hubbard model \cite{hubbard,gutzwiller} and its strong-coupling limit, the $t$-$J$ model \cite{spalek,zhang}.
The Hubbard model is the simplest approximation of the general Hamiltonian of the interacting electron systems, in which all the Coulomb interaction terms except the on-site term ($U$-term) are neglected.  
It is possible that the neglected interaction terms may play important roles in the understanding of the physics of the strongly correlated electron systems. 
That is, some important physics might have been missed in the Hubbard model.
As a result, the $t$-$J$ model, which is a projection of the Hubbard model to the subspace of singly occupied states, may not be sufficient to describe the essential physics of the strong coupling systems.
In this study, we will apply the Gutzwiller projection scheme to a more general model, in which, in addition to the Hubbard on-site term, all of the two-site interaction terms between nearest-neighbor sites are retained in the approximation of the Coulomb potential energy.
Our study shows that this novel model provides a more complete understanding of the properties of high-temperature superconductors.

\section{The Model}

The general Hamiltonian describing the dynamics of electrons in Wannier representation can be expressed as \cite{hubbard,strack,mahan}
\begin{equation}\label{eq:wannier}
\begin{aligned}{}
H &=\sum_{ij,\sigma}T_{ij}c^{\dagger}_{i \sigma}c_{j \sigma}
+\frac{1}{2}\sum_{ijkl,\sigma\sigma^{'}}
\langle ij|\frac{1}{r}|kl\rangle
c^{\dagger}_{i\sigma}
c^{\dagger}_{j\sigma^{'}}
c_{l\sigma^{'}}
c_{k\sigma},
\end{aligned}
\end{equation}
where 
$c^{\dagger}_{i \sigma}$ and $c_{i \sigma}$
are the creation and annihilation operators for an electron with spin $\sigma$ in a Wannier orbital localized at site $i$,
$T_{ij}$ is the Fourier transform of the band energy
$\epsilon_{\textbf{k}}$
\begin{equation}
\begin{aligned}{}
T_{ij} &=\frac{1}{N} \sum_{\textbf{k}} \epsilon_{\textbf{k}} e^{i\textbf{k}\cdot (\textbf{R}_{i}-\textbf{R}_{j})},
\end{aligned}
\end{equation}
and the Wannier representation matrix element is given by
\begin{equation}\label{eq:wannierelement}
\begin{aligned}{}
\langle ij|\frac{1}{r}|kl\rangle =e^2
\int d \textbf{x} d \textbf{x}^{'} 
\frac{\phi^{*}(\textbf{x}-\textbf{R}_{i})
\phi(\textbf{x}-\textbf{R}_{k})
\phi^{*}(\textbf{x}^{'}-\textbf{R}_{j})
\phi(\textbf{x}^{'}-\textbf{R}_{l})}
{|\textbf{x} - \textbf{x}^{'}|},
\end{aligned}
\end{equation}
where $\phi(\textbf{x}-\textbf{R}_{i})$ and
$\phi^*(\textbf{x}-\textbf{R}_{i})$
are the Wannier functions localized around lattice site $i$. 
In the most general cases where
$i \neq j \neq k \neq l$,
the Wannier matrix elements
$\langle ij|\frac{1}{r}|kl\rangle$
are four-center integrals and the corresponding terms in  
the series of the Coulomb interaction in Eq (\ref{eq:wannier}) are the so-called four-site terms.
Since the Wannier function $\phi(\textbf{x}-\textbf{R}_{i})$ goes to zero rapidly when $\textbf{x}$ is away from $\textbf{R}_{i}$, the matrix element (\ref{eq:wannierelement}) is not negligible only when the sites $i,j,k,l$ are close enough so that the overlaps between the Wannier functions are sufficiently large.
Therefore, the Coulomb interaction energy can be conveniently approximated by a number of its leading terms of the series.
The biggest interaction term is the so-called on-site term ($i=j=k=l$)
\begin{equation}\label{eq:0a}
\begin{aligned}{}
 \frac{1}{2}U \sum_{i\sigma} n_{i \sigma}n_{i \bar{\sigma}},
\end{aligned}
\end{equation}
where 
$n_{i \sigma}=c^{\dagger}_{i \sigma}c_{i \sigma}$ and 
$U=\langle ii|\frac{1}{r}|ii\rangle$.
If only this on-site term is taken into account, the Hubbard model is obtained.
Apparently, the next leading terms are the ones where the set $\{i,j,k,l\}$ actually consists of only one pair of nearest-neighbor sites, which may be referred to as the two-site interaction terms \cite{twosite}.
In this work, we take one step further beyond the Hubbard model.
We retain all of the terms up to the two-site interaction terms in the approximation of the Coulomb interaction.
Such a Hamiltonian can be written as
\begin{equation}\label{eq:tpq0}
\begin{aligned}{}
H & \simeq T_{0} \sum_{i\sigma} n_{i\sigma}
-t\sum_{ij,\sigma}\gamma_{ij}c^{\dagger}_{i \sigma}c_{j \sigma}
+ U \sum_{i} n_{i \uparrow}n_{i\downarrow}
\\
&+\frac{1}{2} V \sum_{ij,\sigma\sigma^{'}} \gamma_{ij} n_{i \sigma}n_{j \sigma^{'}}
+X \sum_{ij,\sigma}\gamma_{ij}c^{\dagger}_{i \sigma} c_{j \sigma}
\big(n_{i \bar{\sigma}}+ n_{j \bar{\sigma}}\big)
\\
&+\frac{1}{2} Y \sum_{ij,\sigma} \gamma_{ij} \big(
c^{\dagger}_{i\sigma}
c_{j\sigma}
c^{\dagger}_{i\bar{\sigma}}
c_{j\bar{\sigma}}
+\sum_{\sigma'}
c^{\dagger}_{i\sigma}
c^{\dagger}_{j\sigma'}
c_{i\sigma'}
c_{j\sigma}\big),
\end{aligned}
\end{equation}
where
$\gamma_{ij}=1$ 
for nearest-neighbor sites $i,j$, and $0$ otherwise, which restricts the summation over the nearest-neighbor pairs, and
\begin{equation}\label{eq:x}
\begin{aligned}{}
V & = \langle ij|\frac{1}{r}|ij\rangle,
\\
X&= \langle ii|\frac{1}{r}|ij\rangle=
\langle ii|\frac{1}{r}|ji\rangle=
\langle ij|\frac{1}{r}|ii\rangle=
\langle ji|\frac{1}{r}|ii\rangle,
\\
Y &=\langle ii|\frac{1}{r}|jj\rangle=
\langle ij|\frac{1}{r}|ji\rangle.
\end{aligned}
\end{equation}
In fact, the $X$ and $V$ terms have been discussed in literature previously \cite{strack,tao}, and the matrix element $Y$ has been mentioned in \cite{hubbard}.
Please note that in Eq (\ref{eq:tpq0}) only the constant and nearest-neighbor hopping terms of the kinetic energy are considered for consistency.

Using the Gutzwiller projection operator technique, the Hamiltonian (\ref{eq:tpq0}) can be projected to a subspace of the Hilbert space where only empty and singly occupied sites are allowed \cite{gutzwiller,wilson,onevirtualdstate}.
To the leading order of the perturbation expansion, 
the effective Hamiltonian in the subspace of singly occupied states can be written as
\begin{equation}\label{eq:tpq1}
\begin{aligned}{}
H_s &=
T_{0} \sum_{i\sigma} \tilde{n}_{i\sigma}+
\sum_{ij,\sigma}\gamma_{ij} \Big(-t \, \tilde{c}^{\dagger}_{i \sigma}\tilde{c}_{j \sigma}
- \frac{1}{2} J\, 
c^{\dagger}_{i\sigma}
c^{\dagger}_{j \bar{\sigma}}
c_{i \bar{\sigma}}
c_{j\sigma}
\\
&+ \frac{1}{2} p\, \tilde{n}_{i \sigma} \tilde{n}_{j \sigma}
+ \frac{1}{2} q\, \tilde{n}_{i \sigma}\tilde{n}_{j \bar{\sigma}}\Big),
\end{aligned}
\end{equation}
where 
$\tilde{c}^{\dagger}_{i \sigma}=
c^{\dagger}_{i \sigma}(1-n_{i\bar{\sigma}})$, 
$\tilde{c}_{i \sigma}=
c_{i \sigma}(1-n_{i\bar{\sigma}})$,
$\tilde{n}_{i\sigma}=\sum_{\sigma}
\tilde{c}^{\dagger}_{i \sigma}\tilde{c}_{i \sigma}$,
and
the coefficients are given by the following equations
\begin{equation}\label{}
\begin{aligned}{}
p &=V-Y,
\\
q &=V-2 J_{0}, \,\,
J_{0} =(t-X)^2/U',\,\, U'=U+z'(2V-Y),
\\
J &=p-q,
\end{aligned}
\end{equation}
where $0 \leq z'\leq z$.
When $V$=$X$=$Y$=$0$, Eq (\ref{eq:tpq1}) reduces to the $t$-$J$ model \cite{spalek}.
Rescaling $H_s$ with $t$, this model has only three dimensionless independent material parameters $\bar{T}_0=T_0/t$, $\bar{p}=p/t$, and $\bar{q}=q/t$.   
One may expect that the system will exhibit ferromagnetism when $\bar{q}$ is sufficiently larger than $\bar{p}$. 
On the other hand, the system will exhibit antiferromagnetism when $\bar{p}$ is sufficiently larger than $\bar{q}$.
Our studies show that this is true.
More interestingly, we will show that this model will exhibit superconductivity for some parameter range of $\bar{p} > \bar{q}$.

\section{Green Function Method}

We use the Zubarev Green function technique \cite{hubbard,zubarev} to study the physics of the model established above.
Define the grand canonical Hamiltonian operator \cite{fetter}
\begin{equation}\label{eq:k00}
\begin{aligned}{}
K=H_s-\mu N,
\end{aligned}
\end{equation}
where $N$ is the total number operator, $H_s$ is the Hamiltonian given in Eq (\ref{eq:tpq1}), and $\mu$ is the chemical potential of electrons.
The corresponding grand partition function may be written as
\begin{equation}\label{eq:zg}
\begin{aligned}{}
Z_G= \mbox{Tr} e^{-\beta K}=e^{-\beta \Omega}, \,\, \beta=\frac{1}{k T},
\end{aligned}
\end{equation}
where $k$ is the Boltzmann's constant and $\Omega$ is the thermodynamic potential of the system.
Therefore, the ensemble average of any operator $O$ can be calculated by
\begin{equation}\label{eq:ea}
\begin{aligned}{}
\langle O\rangle= \mbox{Tr} \big[O e^{\beta (\Omega-K) }\big].
\end{aligned}
\end{equation}
Note here that $O$ and $K$ are operators, while $\Omega$ is a $c$-number. The ensemble average $\langle O\rangle$ is assumed to be the measurement value of the observable $O$ in experiments.
The retarded Green function of two Fermion operators $A(t)$ and $B(t')$ in the Heisenberg representation can be defined as
\begin{equation}
\begin{aligned}{}
\ll A(t)|B(t')\gg=
-i \theta(t-t')\langle \{A(t),B(t')\}\rangle,
\end{aligned}
\end{equation}
where 
$\{A,B\}=AB+BA$, the Fermion anticommutation relation, $\theta(t)$ is the usual step function, and $\langle ...\rangle$ means the ensemble average with the grand partition function $Z_G$, defined in Eq (\ref{eq:ea}).
Since 
$\ll  A(t)|B(t')\gg$
is a function of $t-t'$, it is convenient to define the Fourier transform
\begin{equation}
\begin{aligned}{}
\ll  A|B\gg_{\omega}=
\int_{-\infty}^{\infty} d (t-t')
\ll  A(t)|B(t')\gg
e^{i \omega (t-t')}.
\end{aligned}
\end{equation}
It can be shown that the retarded Green function in frequency space satisfies the following equation of motion
\begin{equation}\label{eq:em100}
\begin{aligned}{}
\omega \ll  A|B\gg_{\omega}=
\langle \{A,B\}\rangle
+\ll [A,K]|B\gg_{\omega},
\end{aligned}
\end{equation}
where again $\{A,B\}$ is the anticommutator of operators $A$ and $B$, and $[A,K]$ is the commutator of the operators $A$ and $K$.
The ensemble average $\langle B(t)A(t') \rangle$ can be calculated by 
\begin{equation} \label{eq:ea100}
\begin{aligned}{}
\langle B(t)A(t') \rangle=i
\int_{-\infty}^{\infty} \frac{d \omega}{2 \pi}
\frac{e^{-i\omega(t-t')}}{e^{\beta \omega}+1}\Big[
\ll  A|B\gg_{\omega+i0^{+}}-
\ll  A|B\gg_{\omega-i0^{+}}\Big].
\end{aligned}
\end{equation}
Eqs (\ref{eq:em100})-(\ref{eq:ea100}) are applicable to all subsequent studies of the ferromagnetism, antiferromagnetism, and superconductivity of the model given in Eq (\ref{eq:tpq1}).

Due to the complexity of the model, we only consider the simplest situation in this study: a three-dimensional square lattice system. The two-dimensional case and other complicated factors are left as future work. 

\section{Ferromagnetism}

We first study the ferromagnetism of the model
using the standard Zubarev Green function technique.
For a uniform system, we assume that the ensemble average of the electron number operator 
$\langle n_{l \sigma}\rangle$
is independent of lattice site $l$, but dependent on spin $\sigma$.
Denoting 
$n_{\sigma} =\langle n_{l \sigma}\rangle$,
the average number of electrons per site and magnetization can be defined as
\begin{align}
n &=n_{\uparrow}+n_{\downarrow}, \label{eq:nm1_fm}
\\
m &=n_{\uparrow}-n_{\downarrow}, \label{eq:nm2_fm}
\end{align}
from which, $n_{\sigma}$ can be expressed as
\begin{equation}\label{eq:nsigma}
\begin{aligned}{}
n_{\sigma}=\frac{1}{2} (n+\sigma m),
\end{aligned}
\end{equation}
where $\sigma=1 (-1)$ in the expression for spin subscript $\uparrow$ ($\downarrow$).
Here $n$ is a known quantity when the doping level is known, from which the chemical potential $\mu$ is determined.
For electron-doped materials, $n=1+\delta$, while for hole-doped cases, $n=1-\delta$, where $\delta$ is the doping concentration.
We will focus on the hole-doping only in this study.

The analysis above shows that the central task in the study of the magnetic properties is to calculate the ensemble average of the number operator of electrons, $\langle n_{l \sigma}\rangle$.
According to Eq (\ref{eq:ea100}), $\langle n_{l \sigma}\rangle$ can be evaluated from the Green function 
$\ll c_{l\sigma}|c^{\dagger}_{l'\sigma}\gg$.
The retarded Green function in frequency space satisfies the following equation of motion
\begin{equation}\label{eq:em0}
\begin{aligned}{}
\omega \ll c_{l\sigma}|c^{\dagger}_{l'\sigma}\gg_{\omega}=
\delta_{ll'}
+\ll [c_{l\sigma},K]|c^{\dagger}_{l^{'}\sigma}\gg_{\omega},
\end{aligned}
\end{equation}
where 
$[c_{l\sigma},K]$ is the commutator of the operators $c_{l\sigma}$ and $K$, and $K$ is given in Eq (\ref{eq:k00}).
The Green function 
$\ll c_{l\sigma}|c^{\dagger}_{l'\sigma}\gg_{\omega}$
is a function of $\omega$ and 
$\textbf{R}_{l'}-\textbf{R}_{l}$ for systems with translational invariance.

Carrying out the commutation relation 
$
[c_{l\sigma},K],
$
to the lowest order of the decoupling approximation of the sequence of Green function equations, we obtain
\begin{equation}\label{eq:ftfm}
\begin{aligned}{}
\ll c_{l\sigma}|c^{\dagger}_{l'\sigma}\gg_{\omega}=
\frac{1}{N}\sum_{\textbf{k}} 
e^{i \textbf{k}\cdot (\textbf{R}_{l}-\textbf{R}_{l'})}
\frac{1}{\omega+\mu-z a_{\sigma}-b_{\sigma}\gamma(\textbf{k})},
\end{aligned}
\end{equation}
where $z$ is the coordination number and the $\sigma$ dependent coefficients $a_{\sigma}$ and $b_{\sigma}$ are given by the following equations
\begin{equation}
\begin{aligned}{}
a_{\sigma} &= a_{1\sigma}+\sigma \, m \, a_{2\sigma},
\\
b_{\sigma} &= b_{1\sigma}+\sigma \, m \, b_{2\sigma},
\end{aligned}
\end{equation}
\begin{equation}\label{eq:ab_fm}
\begin{aligned}{}
a_{1\sigma} &= T_0 \delta+
t(1+\delta)+(p+q)\delta \Big[\frac{1}{4} (1-\delta^2)+\Lambda^2_{\bar{\sigma}}+\frac{1}{4} m^2 \Big],
\\
a_{2\sigma} &= -T_0-t+\frac{1}{4}(3p-q)
-(p+q)\Big(\frac{\delta^2}{4}
+\Lambda^2_{\bar{\sigma}}-\frac{1}{4}m^2\Big),
\\
b_{1\sigma} &= t \Big[-\frac{1}{4}(1+\delta)^2
+\Lambda^2_{\bar{\sigma}}
+2\Lambda_{\bar{\sigma}}\Lambda_{\sigma}
-\frac{1}{4}m^2 \Big]
-(p-q)\Lambda_{\bar{\sigma}}
\\
&+ \Big[-\frac{1}{2}(p-q)
-2(p+q)\Big(\frac{1}{4}\delta^2-\Lambda^2_{\bar{\sigma}}
+\frac{1}{4}m^2\Big)\Big] \Lambda_{\sigma},
\\
b_{2\sigma} &= -\frac{1}{2} t (1+\delta)-(p+q)\delta \Lambda_{\sigma},
\end{aligned}
\end{equation}
and the dispersion relation $\gamma(\textbf{k})$ is defined as
\begin{equation}\label{eq:dispersion}
\begin{aligned}{}
\gamma(\textbf{k})=\sum_{\textbf{a}} e^{\textbf{k}\cdot \textbf{a}},
\end{aligned}
\end{equation}
where $\textbf{a}$ is the lattice space vector and the summation is over all the nearest-neighbor sites.
In Eq (\ref{eq:ab_fm}), $\Lambda_{\sigma}$ represents the value of the nonzero formal ensemble average  
$\langle c^{\dagger}_{l'\sigma} c_{l\sigma}\rangle$ for the nearest-neighbor sites $l$ and $l'$.
$\Lambda_{\sigma}$ will be calculated self-consistently \cite{formally}.

Using Eqs (\ref{eq:ea100}) and (\ref{eq:ftfm}), the ensemble average 
$\langle c^{\dagger}_{l'\sigma}c_{l\sigma}\rangle$
can be calculated as follows
\begin{equation} \label{eq:average_fm}
\begin{aligned}{}
\langle c^{\dagger}_{l'\sigma} c_{l\sigma}\rangle=
\frac{1}{N}
\sum_{\textbf{k}}e^{i \textbf{k}\cdot (\textbf{R}_{l}-\textbf{R}_{l'})}
\frac{1}{e^{\beta (E_{\sigma}(\textbf{k})-\mu)}+1},
\end{aligned}
\end{equation}
where the elementary excitations energy spectrum is given by
\begin{equation}\label{eq:disp}
\begin{aligned}{}
E_{\sigma}(\textbf{k})=z a_{\sigma}+b_{\sigma}\gamma(\textbf{k}).
\end{aligned}
\end{equation}
Eq (\ref{eq:average_fm}) is the basic formula for the ensemble average
$\langle c^{\dagger}_{l'\sigma} c_{l\sigma}\rangle$, 
from which we are able to obtain the self-consistent equations of
$n_\sigma$ and $\Lambda_\sigma$
\begin{align}
n_{\sigma} &=z
\int_{-1}^{1} dx \rho(x)
\frac{1}{e^{\beta z b_{\sigma} (x-x_{0\sigma})}+1},
\label{eq:nlam1b_fm}
\\
\Lambda_{\sigma} &=\frac{z^2}{z_{1}}
\int_{-1}^{1} dx \rho(x)
\frac{x}{e^{\beta z b_{\sigma} (x-x_{0\sigma})}+1},
\label{eq:nlam2b_fm}
\end{align}
where 
$x_{0\sigma}=(\bar{\mu}-a_{\sigma})/b_{\sigma}$ with
$\bar{\mu}=\mu/z$,
and the density of state can be expressed as
\begin{equation}\label{eq:rho}
\begin{aligned}{}
\rho(x)=\frac{1}{3\pi} \sqrt{1-x^2},
\end{aligned}
\end{equation}
for three dimensional systems \cite{hubbard}.
Here the summation over $\textbf{k}$ has been converted to an integral over $x\rightarrow \gamma(\textbf{k})/z$ and $z_1=z(1-\delta)$, the average number of the nonempty nearest-neighbor sites of the site $l$ at doping level $\delta$.

Eqs (\ref{eq:nlam1b_fm}) and (\ref{eq:nlam2b_fm}) are the self-consistent equations describing the ferromagnetism of the model described by Eq (\ref{eq:tpq1}).
Five unknowns, 
$n_{\uparrow}$, 
$n_{\downarrow}$,
$\Lambda_{\uparrow}$, $\Lambda_{\downarrow}$, and
$\mu$ need to be determined from the system of equations.
When Eqs (\ref{eq:nlam1b_fm}) and (\ref{eq:nlam2b_fm}) are written down explicitly for each spin, one has four independent equations.
The fifth equation is given in Eq (\ref{eq:nm1_fm}), which indicates that there is only one independent unknown in the set of $n_{\uparrow}$ and 
$n_{\downarrow}$.
Given $\delta$ (or $n=1-\delta$) and $T$ (or $\beta=1/k T)$, solving Eqs (\ref{eq:nm1_fm}), (\ref{eq:nlam1b_fm}), and (\ref{eq:nlam2b_fm}) for $n_{\sigma}$, $\Lambda_\sigma$, and $\mu$, we can obtain the information about the magnetization $m$ as a function of $\delta$ and $T$. 
Of course, $m$ depends on the material parameters
$\bar{T_0}=T_0/t$, $\bar{p}=p/t$, and $\bar{q}=q/t$ as well.

The system can be solved numerically.
We briefly outline the main features of the ferromagnetism as follows:

1. An increase in $\bar{q}$ or decrease in $\bar{p}$ is favorable for the magnetization $m$.\\
2. The magnetization $m$ decreases with temperature.\\
3. The doping dependence of $m$ is complicated. 
The favorability of doping for $m$ depends on the values of $\bar{p}$ and $\bar{q}$.

The detailed results will be presented in a subsequent paper.
Our main purpose of discussing the ferromagnetism here is to examine the singularity of the self-consistent equations and make a comparison with that of the antiferromagnetism and superconductivity.

The most interesting feature of the system of equations (\ref{eq:nlam1b_fm}) and (\ref{eq:nlam2b_fm}) is that, when $m=0$,  there is no singularity as $T \rightarrow 0$ (or $\beta \rightarrow \infty$).
This feature persists even for the situation of $\delta=0$ and $T \rightarrow 0$.
When $\beta \rightarrow \infty$, the integrals in Eq (\ref{eq:nlam1b_fm}) and (\ref{eq:nlam2b_fm}) can be carried out analytically with the following results:
For $b_{\sigma}>0$, integrating gives
\begin{equation}\label{eq:fm0K1}
\begin{aligned}{}
n_{\sigma}&=\frac{1}{2}+\frac{1}{\pi} \Big[x_{0\sigma} \sqrt{1-x^2_{0\sigma}}+\arcsin(x_{0\sigma})\Big],
\\
\Lambda_{\sigma} &=-\frac{4}{\pi z_{1}}(1-x^2_{0\sigma})^{3/2},
\end{aligned}
\end{equation}
while for $b_{\sigma}<0$, we have
\begin{equation}\label{eq:fm0K2}
\begin{aligned}{}
n_{\sigma} &=\frac{1}{2}-\frac{1}{\pi}\Big[x_{0\sigma} \sqrt{1-x^2_{0\sigma}}+\arcsin(x_{0\sigma})\Big],
\\
\Lambda_{\sigma} &=\frac{4}{\pi z_{1}}(1-x^2_{0\sigma})^{3/2},
\end{aligned}
\end{equation}
Indeed, Eqs (\ref{eq:fm0K1}) and (\ref{eq:fm0K2}) are not singular for any value of $m$ even for the case of $\delta=0$.
This means that the model of Eq (\ref{eq:tpq1}) does not necessarily have a ferromagnetic phase with nonzero $m$ at the condition of $T \rightarrow 0$.
That is, $m=0$ is a possible solution of the model 
at zero temperature.
As we will see below, this is in remarkable contrast to the cases of the antiferromagnetism and superconductivity of this model.

\section{Antiferromagnetism}

We now proceed to study the antiferromagnetism of the model.
As we will see below, this model exhibits much more interesting antiferromagnetic features than those of the ferromagnetism.
The most striking one is that the system is necessarily in an antifierrormagnetic state as $T \rightarrow 0$ at $\delta=0$.

The formalism is similar to the discussion of the ferromagnetism.
It is convenient, however, to formally divide the original lattice into two sublattices so that the nearest-neighbor sites belong to different sublattices.
We use $i_1$, $j_1$, ..., to label the sites in the first sublattice, and $i_2$, $j_2$, ..., to label the sites in the second sublattice.
The central task is to calculate the average number of electrons per site for each spin direction in both sublattices.
Denoting the mean number of electrons with spin $\sigma$ in the sublattices as 
$n_{1\sigma}=\langle n_{i_{1}\sigma}\rangle$ and
$n_{2\sigma}=\langle n_{i_{2}\sigma}\rangle$ respectively,
we can define the mean number of electrons per site and magnetization for the first sublattice
\begin{align}\label{eq:n1m1}
n_1 &= n_{1 \uparrow}+n_{1 \downarrow},\\ 
m_1 &= n_{1 \uparrow}-n_{1 \downarrow}.
\end{align}
Similarly, for the second sublattice, we have
\begin{align}
n_2 &= n_{2 \uparrow}+n_{2 \downarrow},\\ 
m_2 &= n_{2 \uparrow}-n_{2 \downarrow}. \label{eq:n2m2}
\end{align}
Eqs (\ref{eq:n1m1})-(\ref{eq:n2m2}) immediately lead to the following expressions for each sublattice
\begin{align}\label{eq:nm1_af}
n_{1\sigma} &= \frac{1}{2}(n_{1}+\sigma m_{1}),\\ 
n_{2\sigma} &= \frac{1}{2}(n_{2}+\sigma m_{2}). \label{eq:nm2_af}
\end{align}
We make two important assumptions: 
1) $n_1=n_2=n$, and
2) $m_1=-m_2=m$. 
Therefore, Eqs (\ref{eq:nm1_af}) and (\ref{eq:nm2_af}) become
\begin{align}\label{eq:nm1_af_b}
n_{1\sigma} &= \frac{1}{2}(n+\sigma m),\\
n_{2\sigma} &= \frac{1}{2}(n-\sigma m), \label{eq:nm2_af_b}
\end{align}
where $n=1-\delta$ and $m$ will be calculated self-consistently.
The above analysis shows that we only need to find the average electron number in one sublattice.
For example, if $n_{1\sigma}$ is known, then
$m=n_{1 \uparrow}-n_{1 \downarrow}$ is known, and therefor 
$n_{2\sigma}$ is known.

Similar to the ferromagnism case, the ensemble averages $n_{1\sigma}=\langle n_{i_{1}\sigma}\rangle$ 
can be calculated from the Green function
$\ll c_{l_{1}\sigma}|c^{\dagger}_{l^{'}_{1}\sigma}\gg$.
Nevertheless, further analysis shows that the two Green functions 
$\ll c_{l_{1}\sigma}|c^{\dagger}_{l^{'}_{1}\sigma}\gg$ and 
$\ll c_{l_{2}\sigma}|c^{\dagger}_{l^{'}_{1}\sigma}\gg$ are coupled to each other even at the lowest order of the decoupling approximation of the Green function equations.
Therefore we need to discuss the two Green functions simultaneously.
In frequency space, the two Green function satisfy the following equations of motion
\begin{equation}\label{eq:em1_af}
\begin{aligned}{}
\omega \ll c_{l_{1}\sigma}|c^{\dagger}_{l^{'}_{1}\sigma}\gg_{\omega}=
\delta_{ll'}
+\ll [c_{l_{1}\sigma},K]|c^{\dagger}_{l^{'}_{1}\sigma}\gg_{\omega},
\end{aligned}
\end{equation}
\begin{equation}\label{eq:em2_af}
\begin{aligned}{}
\omega \ll c_{l_{2}\sigma}|c^{\dagger}_{l^{'}_{1}\sigma}\gg_{\omega}=
\ll [c_{l_{2}\sigma},K]|c^{\dagger}_{l^{'}_{1}\sigma}\gg_{\omega},
\end{aligned}
\end{equation}
where the anticommutation relations
$\{c_{l_{1}\sigma},c^{\dagger}_{l^{'}_{1}}\}=\delta_{ll'}$ and
$\{c_{l_{2}\sigma},c^{\dagger}_{l^{'}_{1}}\}=0$
have been used.
Using Eq (\ref{eq:k00}), the commutation relations 
$[c_{l_{1}\sigma},K]$ and 
$[c_{l_{2}\sigma},K]$
can be obtained exactly.
Nonetheless, an approximation must be used to break off the chain of Green function equations.
To the lowest order of the decoupling approximation, the Green functions can expressed as
\begin{align}\label{eq:ft1_af}
\ll c_{l_{1}\sigma}|c^{\dagger}_{l^{'}_{1}\sigma}\gg_{\omega} &=
\frac{2}{N}
\sum^{N/2}_{\textbf{k}} e^{i \textbf{k}\cdot (\textbf{R}_{l_{1}}-\textbf{R}_{l^{'}_{1}})}
g_{\textbf{k}\sigma}(\omega),
\\ \label{eq:ft2_af}
\ll c_{l_{2}\sigma}|c^{\dagger}_{l^{'}_{1}\sigma}\gg_{\omega} &=
\frac{2}{N}
\sum^{N/2}_{\textbf{k}} e^{i \textbf{k}\cdot (\textbf{R}_{l_{2}}-\textbf{R}_{l^{'}_{1}})}
f_{\textbf{k}\sigma}(\omega),
\end{align}
where 
\begin{align}\label{eq:g_af_b}
g_{\textbf{k}\sigma}(\omega) &=
\frac{A_{\textbf{k}\sigma}}{\omega+\mu-E_{1}(\textbf{k})}+
\frac{B_{\textbf{k}\sigma}}{\omega+\mu-E_{2}(\textbf{k})},
\\ \label{eq:f_af_b}
f_{\textbf{k}\sigma}(\omega) &=
\frac{C_{\textbf{k}}}{\omega+\mu-E_{1}(\textbf{k})}+
\frac{D_{\textbf{k}}}{\omega+\mu-E_{2}(\textbf{k})},
\end{align}
with the elementary excitation spectrum
\begin{align} \label{eq:e1_af}
E_{1}(\textbf{k}) &=z a-\sqrt{\big(z m a'\big)^2+\big(b' \gamma(\textbf{k})\big)^2},
\\  \label{eq:e2_af}
E_{2}(\textbf{k}) &=z a+\sqrt{\big(z m a'\big)^2+\big(b' \gamma(\textbf{k})\big)^2},
\end{align}
and the partial-fraction decomposition coefficients
\begin{align} \label{eq:ak_af}
A_{\textbf{k}\sigma} &=
\frac{1}{2}\Big[1-\frac{\sigma z m a'}{\sqrt{\big(z m a'\big)^2+\big(b' \gamma(\textbf{k})\big)^2}}\Big],
\\ \label{eq:bk_af}
B_{\textbf{k}\sigma} &=
\frac{1}{2}\Big[1+\frac{\sigma z m a'}{\sqrt{\big(z m a'\big)^2+\big(b' \gamma(\textbf{k})\big)^2}}\Big],
\\ \label{eq:ck_af}
C_{\textbf{k}} &=-
\frac{b' \gamma(\textbf{k})}{2 \sqrt{\big(z m a'\big)^2+\big(b' \gamma(\textbf{k})\big)^2}}=D_{\textbf{k}},
\end{align}
where $\gamma(\textbf{k})$ is given in Eq (\ref{eq:dispersion}), $b'=b(1-\delta)$, and the coefficients $a$, $a'$, and $b$ are given the following equations
\begin{align}\label{eq:a_af}
a &= \frac{T_0}{z}\delta+t(1+\delta) \Lambda
+(p+q)\delta \Big[\frac{1}{4}(1-\delta^2)
+\Lambda^2+\frac{1}{4} m^2\Big],
\\\label{eq:ap_af}
a' &= -\frac{T_0}{z}+t \Lambda + \frac{1}{4}(3 q-p)
+(p+q)\Big[-\frac{1}{4}\delta^2+\Lambda^2+\frac{1}{4} m^2\Big],
\\\label{eq:b_af_b}
b &= t\Big[-\frac{1}{4}(1+\delta)^2+3\Lambda^2
+\frac{1}{4}m^2\Big]+\Big[ -J -\frac{1}{2}(p-q)
\nonumber
\\
&+2(p+q)(-\frac{1}{4}\delta^2+\Lambda^2+\frac{1}{4}m^2)\Big] 
\Lambda.
\end{align}
Here $\Lambda$ is the value of the nonzero formal ensemble average
$\langle c^{\dagger}_{l_1\sigma} c_{l_2\sigma}\rangle$, where $l_1$ and $l_2$ are the nearest-neighbor sites.
We have assumed that $\Lambda$ is independent of spin $\sigma$.
Similar to the ferromagnetism case, $\Lambda$ will be calculated self-consistently \cite{formally}.
Please note that $a$, $a'$, and $b$ are all independent of $\sigma$ explicitly.
As can be seen from Eqs (\ref{eq:e1_af}) and (\ref{eq:e2_af}), two branches of the elementary excitation spectrum exist in the system, both of which are spin $\sigma$ independent.
The coefficients $A_{\textbf{k}\sigma}$ and $B_{\textbf{k}\sigma}$ depend on spin $\sigma$ explicitly, while $C_{\textbf{k}}$ and $D_{\textbf{k}}$ are spin independent.
Therefore $f_{\textbf{k}\sigma}(\omega)$ does not depend on $\sigma$ explicitly, which results from the assumption that $\Lambda$ is independent of $\sigma$.

Using Eqs (\ref{eq:ea100}), (\ref{eq:ft1_af}) and (\ref{eq:ft2_af}), we can obtain the following ensemble averages
\begin{align} \label{eq:ea1_af}
\langle c^{\dagger}_{l^{'}_{1}\sigma} c_{l_{1}\sigma}\rangle &=
\frac{2}{N}
\sum^{N/2}_{\textbf{k}} e^{i \textbf{k}\cdot
(\textbf{l}_{1}-\textbf{l}^{'}_{1})}
\Big[
\frac{A_{\textbf{k}\sigma}}{e^{\beta(E_{1}(\textbf{k})-\mu)}+1}+
\frac{B_{\textbf{k}\sigma}}{e^{\beta(E_{2}(\textbf{k})-\mu)}+1}
\Big],
\\ \label{eq:ea2_af}
\langle c^{\dagger}_{l^{'}_{1}\sigma} c_{l_{2}\sigma}\rangle &=
\frac{2}{N}
\sum^{N/2}_{\textbf{k}} e^{i \textbf{k}\cdot
(\textbf{l}_{2}-\textbf{l}^{'}_{1})}
\Big[
\frac{C_{\textbf{k}}}{e^{\beta(E_{1}(\textbf{k})-\mu)}+1}+
\frac{D_{\textbf{k}}}{e^{\beta(E_{2}(\textbf{k})-\mu)}+1}
\Big],
\end{align}
where we have used $\textbf{l}_{1}-\textbf{l}^{'}_{1}$ to represent
$\textbf{R}_{l_{1}}-\textbf{R}_{l^{'}_{1}}$ for conciseness.
Setting $l_1=l'_1$ in Eq (\ref{eq:ea1_af}), we have
\begin{align} \label{eq:n1sigma}
n_{1\sigma} &=
\frac{2}{N}
\sum^{N/2}_{\textbf{k}}
\Big[
\frac{A_{\textbf{k}\sigma}}{e^{\beta(E_{1}(\textbf{k})-\mu)}+1}+
\frac{B_{\textbf{k}\sigma}}{e^{\beta(E_{2}(\textbf{k})-\mu)}+1}
\Big].
\end{align}
Noticing 
$n=n_{1\uparrow}+n_{1\downarrow}$ and $m=n_{1\uparrow}-n_{1\downarrow}$, and using Eqs (\ref{eq:ak_af})-(\ref{eq:bk_af}), Eq (\ref{eq:n1sigma}) immediately leads to the following self-consistent equations for $\mu$ and $m$
\begin{align} \label{eq:n_af}
n &=\frac{2}{N}\sum^{N/2}_{\textbf{k}}
\Big[
\frac{1}{e^{\beta(E_{1}(\textbf{k})-\mu)}+1}+
\frac{1}{e^{\beta(E_{2}(\textbf{k})-\mu)}+1}
\Big],
\end{align}
\begin{align} \label{eq:m_af}
1 &=\frac{2}{N}\sum^{N/2}_{\textbf{k}}
\frac{-z a'}{\sqrt{\big(z m a'\big)^2+\big(b \gamma(\textbf{k})\big)^2}}
\Big[
\frac{1}{e^{\beta(E_{1}(\textbf{k})-\mu)}+1}-
\frac{1}{e^{\beta(E_{2}(\textbf{k})-\mu)}+1}
\Big].
\end{align}
The self-consistent equation for $\Lambda$ can be obtained from Eq (\ref{eq:ea2_af}) by setting $l_2$ and $l'_1$ as the nearest neighbors
\begin{align}\label{eq:lambda_af_b}
z_1 \Lambda &=\frac{2}{N}\sum^{N/2}_{\textbf{k}}
\frac{-b' \gamma(\textbf{k})^2}{2 \sqrt{\big(z m a'\big)^2+\big(b' \gamma(\textbf{k})\big)^2}}
\Big[
\frac{1}{e^{\beta(E_{1}(\textbf{k})-\mu)}+1}
-
\frac{1}{e^{\beta(E_{2}(\textbf{k})-\mu)}+1}
\Big],
\end{align}
where Eq (\ref{eq:ck_af}) has been used.
Here $z_1=z(1-\delta)$, the average number of the nonempty nearest neighbors of the site $l_{2}$.
Eqs (\ref{eq:e1_af}) and (\ref{eq:e2_af}) show that 
$E_{1}(\textbf{k})\leq E_{2}(\textbf{k})$.
Thus from Eqs (\ref{eq:m_af}) and (\ref{eq:lambda_af_b}), we see that for meaningful solutions,  1) $a'$ must be negative, and 2) $\Lambda$ and $b$ must have the opposite sign.

Finally, introducing a new variable
$x\rightarrow \gamma(\textbf{k})/z$, 
and noticing that the integrand depends on $x$ through $x^2$, for the case of $b>0$, the self-consistent equations can be written in the following integration form
\begin{align} \label{eq:sc1_af_c}
n &=2 z \int_{0}^{1} dx \rho(x)
\Big[
\frac{1}{e^{\beta\epsilon_{1}(x)}+1}+
\frac{1}{e^{\beta\epsilon_{1}(x)}+1}
\Big],
\\ \label{eq:sc2_af_c}
\Lambda &=-\frac{z^2}{z_1}\int_{0}^{1} dx \rho(x)
\frac{x^2}{\sqrt{\bar{m}^2+x^2}}
\Big[
\frac{1}{e^{\beta\epsilon_{1}(x)}+1}-
\frac{1}{e^{\beta\epsilon_{1}(x)}+1}
\Big],
\\ \label{eq:sc3_af_c}
b' &=-2z a'\int_{0}^{1} dx \rho(x)
\frac{1}{\sqrt{\bar{m}^2+x^2}}
\Big[
\frac{1}{e^{\beta\epsilon_{1}(x)}+1}-
\frac{1}{e^{\beta\epsilon_{1}(x)}+1}
\Big],
\end{align}
where the density of states $\rho(x)$ is given in Eq (\ref{eq:rho}) and the two branches of energy excitation spectrum take the form
\begin{align} \label{eq:e1_af_c}
\epsilon_{1}(x) &=z b'\Big[-x_0-\sqrt{\bar{m}^2+x^2} \,\,\Big],
\\ \label{eq:e2_af_c}
\epsilon_{2}(x) &=z b'\Big[-x_0+\sqrt{\bar{m}^2+x^2} \,\,\Big],
\end{align}
where 
$x_0=(\mu/z-a)/b'$ and $\bar{m}^2=(m a'/b')^2$.
Eqs (\ref{eq:sc1_af_c})-(\ref{eq:sc3_af_c}) are the basic self-consistent equations for antiferromagnetism of the model in Eq  (\ref{eq:tpq1}).
At a given condition of the doping level $\delta$ and temperature $T$, three unknowns, $x_0$, $\Lambda$, and $m$ can be determined from the system of equations.
Of course, $x_0$, $\Lambda$, and $m$ also depend on the dimensionless material parameters $\bar{T}_0=T_0/t$, $\bar{p}=p/t$, $\bar{q}=q/t$.
In the following, we discuss the case of $T_0=0$.

The system is in ground state at absolute zero temperature.
This is the $\delta$-axis in the $\delta$-$T$ plane.
When $T \rightarrow 0$, the integral in Eq (\ref{eq:sc1_af_c}) can be carried out analytically with the result
\begin{align}\label{eq:sc0_1_af}
\delta &=\frac{2}{\pi}
\big[\xi \sqrt{1-\xi^2}+\arcsin(\xi)\big], 
\end{align}
where $\xi=\sqrt{x_0^2-\bar{m}^2}$.
In the case of $b>0$, $x_0 < 0$ is applicable to the case of $n=1-\delta$ (hole doping), while $x_0 > 0$ is applicable to the case of $n=1+\delta$ (electron doping).
When $T \rightarrow 0$, for the hole doping case ($x_0 < 0$), Eqs (\ref{eq:sc2_af_c}) and (\ref{eq:sc3_af_c}) can be written as
\begin{align}\label{eq:sc0_2_af}
\Lambda &=-\frac{z^2}{z_1}\int_{\xi}^{1} dx \rho(x)
\frac{x^2}{\sqrt{\bar{m}^2+x^2}},
\\\label{eq:sc0_3_af}
b' &=-2za' \int_{\xi}^{1} dx \rho(x)
\frac{1}{\sqrt{\bar{m}^2+x^2}},
\end{align}
where $\xi$ is given by Eq (\ref{eq:sc0_1_af}).
There is no singularity as long as either $\xi$ or $\bar{m}$ is nonzero.
The quantity $\xi$ depends on $\delta$ only and Eq (\ref{eq:sc0_1_af}) is not coupled with Eqs (\ref{eq:sc0_2_af}) and (\ref{eq:sc0_3_af}).
Thus $\xi$ can be solved separately from Eq (\ref{eq:sc0_1_af}) for a given $\delta$.
When $\delta$ is small, to the leading order, Eq (\ref{eq:sc0_1_af}) can be solved with the result
\begin{align}
\xi \simeq \frac{\pi}{4}\delta,
\end{align}
which leads to an approximate relation between $x_0$ and $\bar{m}$ on the
$\delta$-axis ($T\rightarrow 0$).
\begin{align}\label{eq:x0mba_af}
x_0^2 \simeq \big(\frac{\pi}{4}\delta\big)^2+\bar{m}^2.
\end{align}
A more complete solution of the system of equations (\ref{eq:sc0_1_af})-(\ref{eq:sc0_3_af}) can be obtained numerically for given $\delta$, $\bar{p}$, and $\bar{q}$.
Now we discuss two special points on the $\delta$-axis ($T\rightarrow 0$).

An interesting point on the $\delta$-axis is where $m$ vanishes.
At this point, $\xi=|x_0|$ and the integrals in the self-consistent equations (\ref{eq:sc0_2_af})-(\ref{eq:sc0_3_af}) can be carried out analytically.
Use $\delta_0$ to denote the doping concentration at this point, which can be determined by the following equations 
\begin{align}\label{eq:sc0_d01_af}
\delta_0 &=-\frac{2}{\pi}
\Big[x_0 \sqrt{1-x_0^2}+\arcsin(x_0)\Big], 
\\ \label{eq:sc0_d02_af}
\Lambda &= -\frac{4}{\pi z_1}\Big(1-x_0^2\Big)^{3/2},
\\ \label{eq:sc0_d03_af}
b' &=\frac{4}{\pi} a' \Big[
\sqrt{1-x_0^2}+\ln{|x_0|}-
\ln{\Big(1+\sqrt{1-x_0^2}\Big)} \,\Big].
\end{align}
To the leading order of $\delta_0$, $x_0$ can be solved from Eq (\ref{eq:sc0_d01_af}) as
\begin{align}
x_0 \simeq -\frac{\pi}{4} \delta_0.
\end{align}
The term $\ln{|x_0|}$ in Eq (\ref{eq:sc0_d03_af}) shows that a meaningful $\delta_0$ must be finite.
Solving $x_0$ and $\Lambda$ from Eqs (\ref{eq:sc0_d01_af}) and (\ref{eq:sc0_d02_af}), and then substituting their values into Eq (\ref{eq:sc0_d03_af}), we obtain a single variable equation for $\delta_0$.
Given $\bar{p}$ and $\bar{q}$, this equation can be solved to obtain $\delta_0 (\bar{p},\bar{q})$.
In practice, the value of $\delta_0$ is known from experiments, so that $\delta_0 (\bar{p},\bar{q})$ specifies a relation between the dimensionless parameters $\bar{p}$ and $\bar{q}$.
For example, in the case of $b>0$, for $\delta_0=0.05$, $\bar{q}=0.63$ corresponds to $r=\bar{p}/\bar{q} \simeq 2.5$, while $\bar{q}=1$ corresponds to $r=\bar{p}/\bar{q} \simeq 3.73$.

The most interesting case is the ground state at half-filling.
This is the origin of the $\delta$-$T$ plane.
Eq (\ref{eq:sc0_1_af}) indicates that $\xi=0$ and therefore $\tilde{x}_0^2=\bar{m}^2$ at this point.
Noticing that at 
$\delta=0$, $z_1=z$, and $b'=b$,
the self-consistent equations become
\begin{align}\label{eq:sc00_2_af}
\Lambda &=-z\int_{0}^{1} dx \rho(x)
\frac{x^2}{\sqrt{\bar{m}^2+x^2}},
\\ \label{eq:sc00_3_af}
b &=-2za' \int_{0}^{1} dx \rho(x)
\frac{1}{\sqrt{\bar{m}^2+x^2}}.
\end{align}
Apparently, the integral in Eq (\ref{eq:sc00_3_af}) is divergent when $m=0$, which indicates that $m=0$ is not a possible solution of the system as 
$T \rightarrow 0$ at $\delta=0$.
That is, the system must be in an antiferromagnetic state with a nonzero $m$ at the origin of the $\delta$-$T$ plane.
This is one of the most striking features of the model described by Eq (\ref{eq:tpq1}).
Since in this case $T\rightarrow 0$ and $\delta=0$, the magnetization $m$ depends on the two parameters $\bar{q}$ and $\bar{p}$ (or $\bar{q}$ and $r=\bar{p}/\bar{q}=p/q$ ) only.
Our results show that for a given value of $\bar{q}$, a nonzero $m$ starts with a threshold value $r \ge r_0$.
Then $m$ increases with $r$, and finally saturates to a maximum value of $m_{max} \simeq 0.6698$ when $r$ is sufficiently large.
Generally, the greater the $\bar{q}$, the greater the threshold value $r_0$. 
The threshold value of $r$ for a given $\bar{q}$ is related to the condition of $a'<0$ and $b'>0$. 

\begin{figure}[!h]
\centering
\includegraphics[clip,width=10cm, height=10cm]{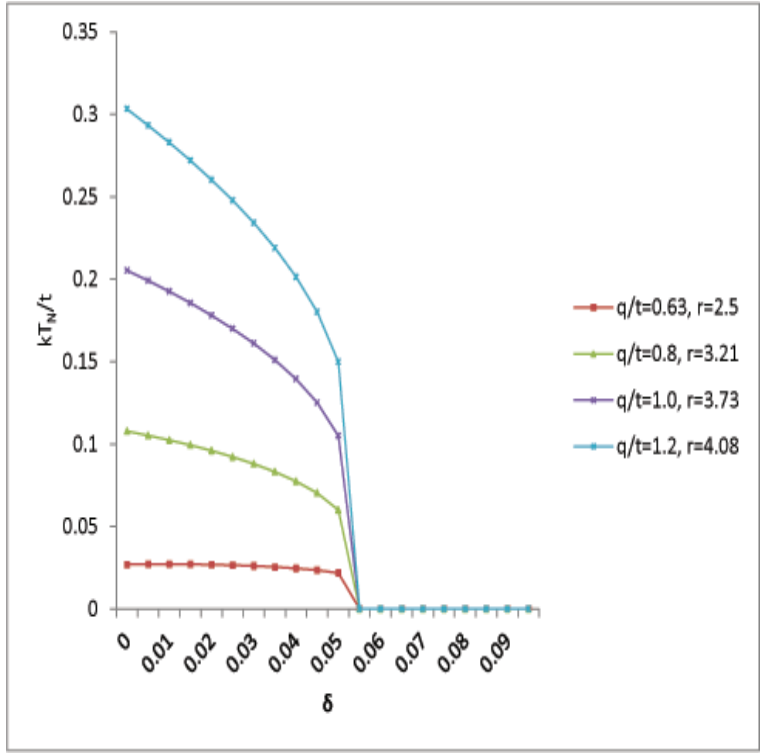}
\caption{Antiferromagnetic phase boundary lines for typical values of $\bar{q}=q/t$ and $r$. The region below each line is the antiferromagnetic state.}
\end{figure}  
 
In the simplest case where $m=0$, for the case of $b>0$, the self-consistent equations become
\begin{align} \label{eq:sc0m_1_af}
n &=2 z \int_{0}^{1} dx \rho(x)
\Big[
\frac{1}{e^{\beta\epsilon_{1}(x)}+1}+
\frac{1}{e^{\beta\epsilon_{1}(x)}+1}
\Big],
\\ \label{eq:sc0m_2_af}
\Lambda &=-\frac{z^2}{z_1}\int_{0}^{1} dx \rho(x) x
\Big[
\frac{1}{e^{\beta\epsilon_{1}(x)}+1}-
\frac{1}{e^{\beta\epsilon_{1}(x)}+1}
\Big],
\\ \label{eq:sc0m_3_af}
b' &=-2z a'\int_{0}^{1} dx \rho(x)
\frac{1}{x}
\Big[
\frac{1}{e^{\beta\epsilon_{1}(x)}+1}-
\frac{1}{e^{\beta\epsilon_{1}(x)}+1}
\Big],
\end{align}
where 
\begin{align}
\epsilon_{1}(x) &=z b' \big(-x_0-x \,\,\big),
\\
\epsilon_{2}(x) &=z b' \big(-x_0+x \,\,\big).
\end{align}
Now there are only two unknowns $x_0$ and $\Lambda$, which can be determined from Eqs (\ref{eq:sc0m_1_af}) and (\ref{eq:sc0m_2_af}) when $\delta$ and $T$ are given.
The third equation (\ref{eq:sc0m_3_af}) will give the N$\acute{e}$el temperature at a doping concentration $\delta$, $T_N(\delta)$, in the $\delta$-$T$ plane, which is the boundary of the area where $m \neq 0$.
Fig. $1$ shows the antiferromagnetic phase diagrams for some typical values of $\bar{q}$ and $r$. 
As discussed above, Eq (\ref{eq:sc0m_3_af}) is singular at the point of $T\rightarrow 0$ and $\delta=0$.
Therefore, the boundary of $m \neq 0$ must bypass the origin of the $\delta$-$T$ plane. 

\section{Superconductivity}

In order to study the superconductivity of the model described by Eq (\ref{eq:tpq1}), one needs to calculate the ensemble average
$
\langle
c_{l\downarrow}
c_{l'\uparrow}
\rangle
$
or its complex conjugate
$
\langle
c^{\dagger}_{l'\uparrow}
c^{\dagger}_{l\downarrow}
\rangle.
$
These quantities provide the information about the state of the Cooper pairs, from which the properties of the superconducting and pseudogap state can be obtained.
The ensemble average 
$
\langle
c^{\dagger}_{l'\uparrow}
c^{\dagger}_{l\downarrow}
\rangle
$
can be calculated from the Green function
$
\ll c^{\dagger}_{l\downarrow}|
c^{\dagger}_{l'\uparrow}\gg.
$
Further analysis shows that the Green function 
$
\ll c^{\dagger}_{l\downarrow}|
c^{\dagger}_{l'\uparrow}\gg
$
is coupled with
$
\ll c_{l\uparrow}|c^{\dagger}_{l'\uparrow}\gg
$
even at the lowest order of decoupling approximation.
Therefore, we need to deal with both
$
\ll c^{\dagger}_{l\downarrow}|
c^{\dagger}_{l'\uparrow}\gg
$
and
$
\ll c_{l\uparrow}|
c^{\dagger}_{l'\uparrow}\gg
$
simultaneously in the study of the superconductivity.
In $\omega$-space, the equations of motion for both Green functions
$
\ll c_{l\uparrow}|c^{\dagger}_{l'\uparrow}\gg_\omega
$
and
$
\ll c^{\dagger}_{l\downarrow}|c^{\dagger}_{l'\uparrow}\gg_\omega
$
are
\begin{equation}\label{eq:em1}
\begin{aligned}{}
\omega \ll c_{l\uparrow}|c^{\dagger}_{l'\uparrow}\gg_{\omega}=
\delta_{ll'}
+\ll [c_{l\uparrow},K]|c^{\dagger}_{l^{'}\uparrow}\gg_{\omega},
\end{aligned}
\end{equation}
\begin{equation}\label{eq:em2}
\begin{aligned}{}
\omega \ll c^{\dagger}_{l\downarrow}|
c^{\dagger}_{l'\uparrow}\gg_{\omega}=
\ll[c^{\dagger}_{l\downarrow},K]|
c^{\dagger}_{l^{'}\uparrow}\gg_{\omega},
\end{aligned}
\end{equation}
where
the Fermion anticommutation relations 
$
\{c_{l\uparrow},
c^{\dagger}_{l'\uparrow}\}=\delta_{ll'}
$
and
$
\{c^{\dagger}_{l\downarrow},
c^{\dagger}_{l'\uparrow}\}=0
$
have been taken into account.
For translational invariant systems,
$
\ll c_{l\uparrow}|c^{\dagger}_{l'\uparrow}\gg_\omega
$
and
$
\ll c^{\dagger}_{l\downarrow}|c^{\dagger}_{l'\uparrow}\gg_\omega
$
are functions of frequency $\omega$ and
$\textbf{R}_{l}-\textbf{R}_{l'}$.

To the lowest order of decoupling approximation, we obtain the Green functions as follows
\begin{equation}\label{eq:ft1_sc}
\begin{aligned}{}
\ll c_{l\uparrow}|
c^{\dagger}_{l'\uparrow}\gg_{\omega}=
\frac{1}{N}
\sum_{\textbf{k}} e^{i \textbf{k}\cdot (\textbf{R}_{l}-\textbf{R}_{l'})}
\Big[\frac{A_{\textbf{k}}}{\omega-E_{\textbf{k}}}+
\frac{B_{\textbf{k}}}{\omega+E_{\textbf{k}}}\Big],
\end{aligned}
\end{equation}
\begin{equation}\label{eq:ft2_sc}
\begin{aligned}{}
\ll c^{\dagger}_{l\downarrow}|
c^{\dagger}_{l'\uparrow}\gg_{\omega}=
\frac{1}{N}
\sum_{\textbf{k}} e^{i \textbf{k}\cdot (\textbf{R}_{l}-\textbf{R}_{l'})}
\Big[\frac{C_{\textbf{k}}}{\omega-E_{\textbf{k}}}+
\frac{D_{\textbf{k}}}{\omega+E_{\textbf{k}}}\Big],
\end{aligned}
\end{equation}
where the partial-fraction decomposition coefficients are given by
\begin{equation}\label{eq:ak_sc}
\begin{aligned}{}
A_{\textbf{k}}=\frac{1}{2}\Big(1+
\frac{-\mu+z a+ b'\gamma_{\textbf{k}}}{E_{\textbf{k}}}\Big),
\end{aligned}
\end{equation}
\begin{equation}\label{eq:bk_sc}
\begin{aligned}{}
B_{\textbf{k}}=\frac{1}{2}\Big(1-
\frac{-\mu+z a+ b'\gamma_{\textbf{k}}}{E_{\textbf{k}}}\Big),
\end{aligned}
\end{equation}
\begin{equation}\label{eq:ck_sc}
\begin{aligned}{}
C_{\textbf{k}}=\frac{1}{2}
\frac{(z c+d'\gamma_{\textbf{k}})\Delta^{\ast}}{E_{\textbf{k}}}
=-D_{\textbf{k}},
\end{aligned}
\end{equation}
and the elementary excitation energy spectrum takes the form
\begin{equation}\label{eq:e_sc}
\begin{aligned}{}
E_{\textbf{k}}=\sqrt{(\mu-z a- b^{'}\gamma_{\textbf{k}})^2 
+(z c+d^{'}\gamma_{\textbf{k}})^2 |\Delta|^2}.
\end{aligned}
\end{equation}
Here the dispersion relation $\gamma_{\textbf{k}}$ is given in Eq (\ref{eq:dispersion}) and
$b'=b(1-\delta)/(1+r)$, $d'=d(1-\delta)r/(1+r)$, where $r=p/q$, and the coefficients $a$, $b$, $c$, and $d$ are expressions of $\Lambda$ and $\Delta$
\begin{equation}\label{eq:a_sc}
\begin{aligned}{}
a= t(1+\delta)\Lambda+(p+q)\delta \Big[\frac{1}{4} (1-\delta^2)+\Lambda^2- |\Delta|^2\Big],
\end{aligned}
\end{equation}
\begin{equation}\label{eq:b_sc}
\begin{aligned}{}
b &= t\Big[-\frac{1}{4}(1+\delta)^2+3\Lambda^2+|\Delta|^2\Big]
+\Big[-\frac{3}{2}(p-q)\\
&+2(p+q)(-\frac{1}{4} \delta^2+\Lambda^2+ |\Delta|^2)\Big]\Lambda,
\end{aligned}
\end{equation}
\begin{equation} \label{eq:c_sc}
\begin{aligned}{}
c= t(1+\delta)+2(p+q)\delta\Lambda,
\end{aligned}
\end{equation}
\begin{equation}\label{eq:d_sc}
\begin{aligned}{}
d= t 2\Lambda-\frac{3}{2}(p-q)+2(p+q)\Big[\frac{1}{4} \delta^2+\Lambda^2+ |\Delta|^2\Big].
\end{aligned}
\end{equation}
It is easy to check that the dimensionless coefficients $\bar{a}=a/t$, $\bar{b}=b/t$, $\bar{c}=c/t$, and $\bar{d}=d/t$  
satisfy the following relations
\begin{align}
\bar{a} &= \bar{c} \Lambda +(\bar{p}+\bar{q})\delta
\Big[\frac{1}{4}(1-\delta^2)-\Lambda^2- |\Delta|^2\Big], 
\\
\bar{b} &= \bar{d} \Lambda -\frac{1}{2}\bar{c}\delta
-\frac{1}{4}(1+\delta)^2+\Lambda^2+|\Delta|^2,
\end{align}
where $\bar{p}=p/t$ and $\bar{q}=q/t$.
Here $\Lambda$ and $\Delta$ are the values of the nonzero formal ensemble averages
$\langle c^{\dagger}_{l\sigma} c_{l'\sigma} \rangle$ and
$\langle c_{l\downarrow} c_{l'\uparrow} \rangle$ 
for the nearest-neighbor sites $l$ and $l'$, which will be calculated self-consistently \cite{formally}.
We have assumed that $\Lambda$ and $\Delta$ do not depend on spin $\sigma$ explicitly in the discussion of superconductivity.

Using Eq (\ref{eq:ea100}), the ensemble averages
$
\langle
c^{\dagger}_{l'\uparrow}
c_{l \uparrow}
\rangle
$
and
$
\langle
c^{\dagger}_{l'\uparrow}
c^{\dagger}_{l \downarrow}
\rangle
$
can be calculated by the following equations
\begin{equation}\label{eq:ea1}
\begin{aligned}{}
\langle c^{\dagger}_{l'\uparrow} c_{l\uparrow}\rangle=
\frac{1}{2}\delta_{ll'}-\frac{1}{2N}
\sum_{\textbf{k}}e^{i \textbf{k}\cdot
(\textbf{l}-\textbf{l'})}
\frac{-\mu+z a+ b'\gamma_{\textbf{k}}}{E_{\textbf{k}}}
\tanh\big(\frac{1}{2}\beta E_{\textbf{k}}\big), 
\end{aligned}
\end{equation}
\begin{equation}\label{eq:ea2}
\begin{aligned}{}
\langle 
c^{\dagger}_{l'\uparrow} 
c^{\dagger}_{l\downarrow}
\rangle=-\frac{\Delta^{\ast}}{2N}
\sum_{\textbf{k}}e^{i \textbf{k}\cdot
(\textbf{l}-\textbf{l'})}
\frac{z c+d'\gamma_{\textbf{k}}}{E_{\textbf{k}}}
\tanh\big(\frac{1}{2}\beta E_{\textbf{k}}\big).
\end{aligned}
\end{equation}
These are the general formulas of the ensemble averages for the study of superconductivity.
Eqs (\ref{eq:ea1}) and (\ref{eq:ea2}) confirm that
$
\langle c^{\dagger}_{l'\uparrow} c_{l\uparrow}\rangle,
$
and
$
\langle c^{\dagger}_{l'\uparrow} c^{\dagger}_{l\downarrow}\rangle
$ 
are functions of $\textbf{R}_{l}-\textbf{R}_{l'}$, which is a result of the translational invariance of the system.

Three quantities $\mu$, $\Lambda$, and $\Delta$ need to be determined from this formalism.
The self-consistent equations for $\mu$, $\Lambda$, and $\Delta$ can be obtained from Eqs (\ref{eq:ea1}) and (\ref{eq:ea2}) with the following form
\begin{align}
\delta&=\frac{1}{N}\sum_{\textbf{k}}
\frac{-\mu+z a+ b'\gamma_{\textbf{k}}}{E_{\textbf{k}}}
\tanh\big(\frac{1}{2}\beta E_{\textbf{k}}\big).
\\
2z_{1} \Lambda&=-\frac{1}{N}\sum_{\textbf{k}}
\frac{(-\mu+z a+ b'\gamma_{\textbf{k}})\gamma_{\textbf{k}}}
{E_{\textbf{k}}}
\tanh\big(\frac{1}{2}\beta E_{\textbf{k}}\big), 
\\
2z_{2} &=-\frac{1}{N}\sum_{\textbf{k}}
\frac{(z c+d'\gamma_{\textbf{k}})\gamma_{\textbf{k}}}
{E_{\textbf{k}}}
\tanh\big(\frac{1}{2}\beta E_{\textbf{k}}\big), 
\end{align}
where 
$z_1=z (1-\delta)/(1+r)$ and $z_2=z (1-\delta)r/(1+r)$,
the approximate average numbers of the nearest neighbors of the site $l$ with the same and opposite spin, respectively.

Furthermore, introducing a new variable 
$x\rightarrow \gamma_{\textbf{k}}/z$, 
the self-consistent equations can be written in the integration form
\begin{align}\label{eq:se1_d}
\delta &=z b'\int_{-1}^{1}d x \rho(x)
\frac{x-x_{0}}{E(x)}
\tanh \Big(\frac{1}{2}z\beta E(x)\Big),
\\\label{eq:se2_d}
2 z_{1}\Lambda &=-z^2 b'\int_{-1}^{1}d x \rho(x)
\frac{(x-x_{0})x}{E(x)}
\tanh \Big(\frac{1}{2}z \beta E(x)\Big), 
\\\label{eq:se3_d}
2 z_{2} &=-z^2 d' \int_{-1}^{1}dx \rho(x)
\frac{(x-x'_{0})x}{E(x)}
\tanh \Big(\frac{1}{2}z \beta E(x)\Big), 
\end{align}
where
$x_0=(\mu/z-a)/b'$, 
$x'_0=-c/d'$, 
the density of states $\rho(x)$ is given in Eq (\ref{eq:rho}), and the elementary excitation energy can be written as
\begin{equation}\label{eq:e_sc_d}
\begin{aligned}{}
E(x)=\sqrt{b^{'2} (x-x_0)^2 +d^{'2}(x-x'_0)^2 \Delta^2}.
\end{aligned}
\end{equation}
In the dimensionless system where the coefficients $a$, $b$, $c$, and $d$ are rescaled with $t$, there are only two independent parameters $\bar{p}$ and $\bar{q}$.
Except for the expression of density of state (\ref{eq:rho}), all equations are applicable to any dimensional systems.

\begin{figure}[!h]
\centering
\includegraphics[clip,width=10cm, height=7cm]{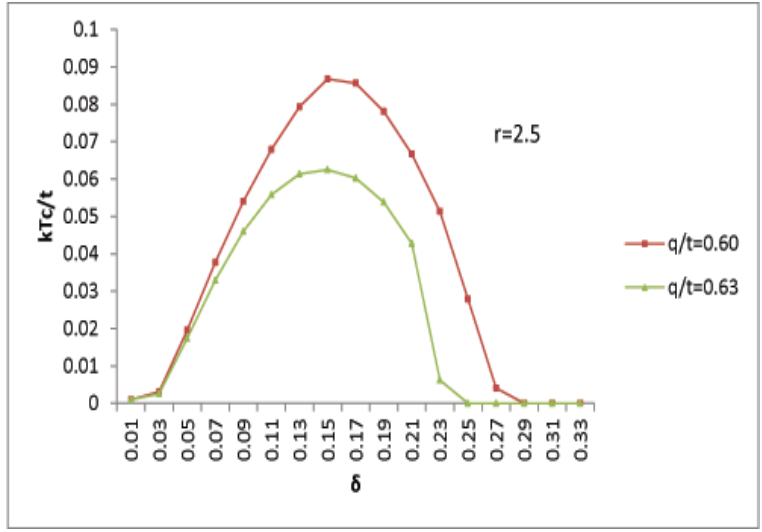}
\caption{Phase diagram of superconductivity state for typical values of $\bar{q}=q/t$ and $r$. The region below each phase boundary line (within each dome) is the superconductivity state.}
\end{figure}  

Eqs (\ref{eq:se1_d})-(\ref{eq:se3_d}) are the basic self-consistent equations for the superconductivity state of the model.
There are three unknowns, $x_0$, $\Lambda$, and $\Delta$, to be determined from the system of equations. 
The most interesting quantity is the energy gap parameter $\Delta$.
It is the order parameter of the superconducting state.
The quantity $x_0$, the effective chemical potential of the system, is relevant to the features of the Fermi surface.
In the case where $d \neq 0$ and $x_0 \neq x'_0$, a nonzero $\Delta$ causes an energy gap at the Fermi surface, as shown in Eq (\ref{eq:e_sc_d}).
This is the most remarkable feature of the superconducting state.
The system of equations (\ref{eq:se1_d})-(\ref{eq:se3_d}) can be solved numerically.
The phase diagram of the superconductivity for some typical values of $\bar{q}$ and $r$ is shown in Fig. 2.
Other detailed results, such as the $\delta$ and $T$ dependence of the energy gap parameter $\Delta$, will be presented in the forthcoming papers.

\section{Pseudogap State}

We believe that the pseudogap state observed in experiments can be described by the special case of the system of equations (\ref{eq:se1_d})-(\ref{eq:se3_d}) where $x_0=x'_0$.
It can be shown that $b = \Lambda d$ in this case, and therefore the self-consistent equations take the following form
\begin{align}\label{eq:se1_pg_b}
\delta &=\frac{z r_1 \Lambda}{\kappa} \int_{-1}^{1}d x \rho(x)
\tanh\Big(\frac{1}{2}z\beta \kappa d (x-x_0)\Big), 
\\\label{eq:se2_pg_b}
2 &=-z \frac{1}{\kappa} \int_{-1}^{1}d x \rho(x) x
\tanh \Big(\frac{1}{2}z \beta \kappa d (x-x_0)\Big), 
\end{align}
\begin{equation}\label{eq:lamdelta}
\begin{aligned}{}
\Lambda^2+\Delta^2 =\frac{1}{4}(1+\delta)^2
+(\bar{p}+\bar{q})\delta^2\Lambda,
\end{aligned}
\end{equation}
where $\kappa$ is given by
\begin{equation}
\begin{aligned}{}
\kappa =\sqrt{(r_1 \Lambda)^2 + (r_2 \Delta)^2}.
\end{aligned}
\end{equation}
Note that there is no singularity in Eqs (\ref{eq:se1_pg_b})-(\ref{eq:se2_pg_b}) at any doping level as $T \rightarrow 0$.
At a given temperature $T$ and doping level $\delta$, the quantities $x_0$, $\Lambda$, and $\Delta$ are determined from Eqs (\ref{eq:se1_pg_b})-(\ref{eq:lamdelta}).

In the pseudogap state, a new degree of freedom, $n^\ast$, which is the number of electrons that cannot `see' the energy gap, emerges.
This is the density of the charge carriers contributing to the electrical conductivity in the pseudogap region.
Explicitly, the condition of $x_0=x'_0$ can expressed as
\begin{equation}\label{eq:cdx0}
\begin{aligned}{}
\bar{c}+x_0 r_2 \bar{d}=0,
\end{aligned}
\end{equation}
where 
\begin{align}\label{eq:cba_pg}
\bar{c}&=2-n+2(\bar{p}+\bar{q})(1-n)\Lambda,
\\ \label{eq:dba_pg}
\bar{d}&=2\Lambda-\bar{p}+2\bar{q}
+2(\bar{p}+\bar{q})\Big[-\frac{1}{4}n(2-n)
+\Lambda^2+\Delta^2\Big].
\end{align}
When $n$ is the total number of the electrons, $n=1-\delta$, Eq (\ref{eq:cdx0}) describes the line of $T^\ast(\delta)$ \cite{tstarline}.
If we require the condition (\ref{eq:cdx0}) to be satisfied for $T<T^\ast$, then the only possibility is that $n<1-\delta$.
We assume that the new degree of freedom $n^\ast$ is determined by Eq (\ref{eq:cdx0}) when $T<T^{\ast}$.
That is, the condition $x_0=x'_0$ determines the number of carries $n^{\ast}$ in the pseudogap state.

\begin{figure}[!h]
\centering
\includegraphics[clip,width=10cm,height=9cm]{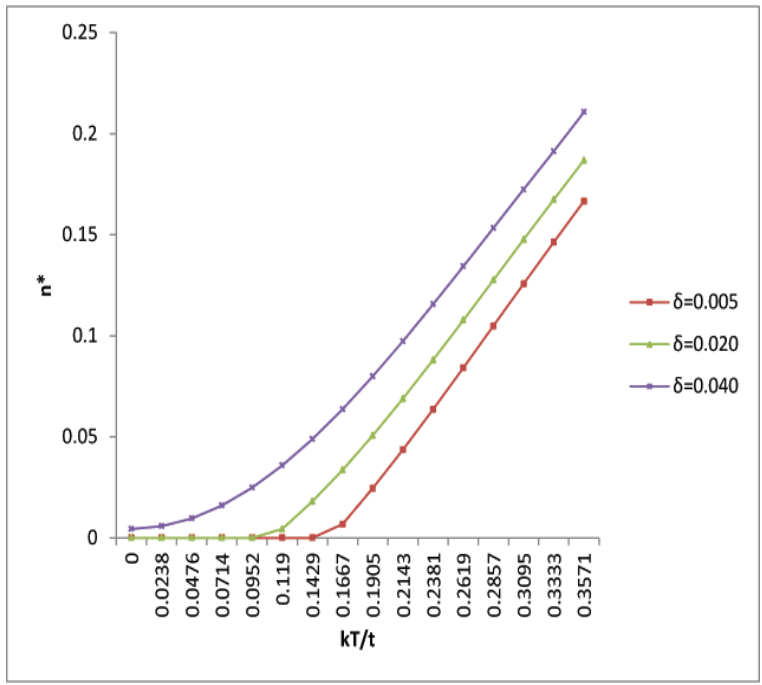}
\caption{Temperature dependence of the charge carriers density $n^{\ast}$ in pseudogap state for typical doping levels ($\bar{q}=0.63$, $r=2.5$).}
\end{figure}  

Replacing $n$ by $n^{\ast}$ in Eq (\ref{eq:cdx0}) leads to a quadratic equation of $n^{\ast}$, which can be solved with the result
\begin{equation}
\begin{aligned}{}
n^\ast &=1-\delta^\ast,
\\
\delta^\ast &=\frac{-1}{\big(\bar{p}+\bar{q}\big) r_2x_0}
\Big[1+2\big(\bar{p}+\bar{q}\big)\Lambda+ \sqrt{D}\Big],
\\
D &=\Big[1+2\big(\bar{p}+\bar{q}\big)\Lambda\Big]^2-2\big(\bar{p}+\bar{q}\big) r_2 x_0
\Big[1+2\Big(-\frac{3}{4}\big(\bar{p}-\bar{q}\big)
\\
&+\Lambda+ \big(\bar{p}+\bar{q}\big)\big(\Lambda^2+\Delta^2\big)\Big)r_2 x_0\Big].
\end{aligned}
\end{equation}
The symbol $\delta^\ast$ should not be confused with the doping parameter.
As Fig. 3 shows, at a doping level $\delta$, $n^\ast$ increases linearly with temperature $T$ for the range of higher $T$.
This remarkable result indicates that the Hall coefficient 
$R_H = 1/e n^\ast$ has a temperature dependence of $1/T$.

As expected, the condition of $n^*=0$ should give the metal-insulator (MI) transition line. 
A semi-phenomenological analysis can show that the resistivity in the pseudogap region
$\rho^{\ast} \sim -1/x_0$ at the Fermi surface.
Detailed discussion of the resistivity and the phase diagram of the pseudogap state (the MI transition line and $T^*$ line) will be presented in forthcoming papers.
 
\section{Conclusion}

We have established a novel model Hamiltonian for the strongly correlated electron systems, which contains the Coulomb potential energy terms up to the order of two-site interactions.
The magnetic properties (ferromagnetism and antiferromagnetism) of this model have been studied.
Our results indicate that the system is not necessarily in the ferromagnetic state as $T\rightarrow 0$ at any doping level.
Nevertheless, the system must be in an antiferromagnetic state with a nonzero magnetization at the state point of $T\rightarrow 0$ and $\delta=0$.
Moreover, the system exhibits a superconducting state in a doped region at sufficiently low temperatures.
The phase diagrams for antiferromagnetism and superconductivity of the model are presented.
This model also predicts the temperature dependence of $1/T$ of the Hall coefficient in the pseudogap state.


\begin{thebibliography}{9}

\bibitem{bednorz}
J. G. Bednorz and K. A. M\"uller, Z. Phys. B \textbf{64}, 189 (1986).

\bibitem{hubbard}
J. Hubbard, Proc. R. Soc. London \textbf{A 276}, 238 (1963), Proc. R. Soc. London \textbf{A 281}, 401 (1964).

\bibitem{gutzwiller}
M. C. Gutzwiller, Phys. Rev. Lett. \textbf{10}, 159 (1963).

\bibitem{spalek}
https://arxiv.org/pdf/0706.4236v1.pdf
J. Spalek, \textit{t-J model then and now: A personal perspective from the pioneering times.}

\bibitem{zhang}
F. C. Zhang and T. M. Rice, Phys. Rev. B \textbf{37}, 3759 (1988).

\bibitem{strack}
R. Strack and D. Vollhardt, Phys. Rev. Lett. \textbf{70}, 2637 (1993)

\bibitem{mahan}
Gerald D. Mahan, \textit{Many-Particle Physics} (Plenum Press, New York, 1990)

\bibitem{twosite}
From now on, we will use the term \textit{two-site interaction terms} to mean 1) the interaction term is a two-center integral and 2) the two sites involved are the nearest neighbors.

\bibitem{tao}
T. Sun, Phil. Mag. Lett. \textbf{73}, 201 (1996)

\bibitem{wilson}
 I. Hubac and S. Wilson \textit{Brillouin-Wigner Methods for Many-Body Systems} (Springer Netherlands, 2010)

\bibitem{onevirtualdstate}
D. Vollhardt, in \textit{Proceedings of the International School of Physics} "Enrico Fermi" Course CXXI, edited by R. A. Broglia and J. R. Schrieffer (North Holland, Amsterdam, 1994), p. 31.

\bibitem{zubarev}
D. N. Zubarev, Sov. Phys. Uspekhi \textbf{3}, 320 (1960).

\bibitem{fetter}
A. L. Fetter and J. D. Walecka, \textit{Quantum Theory of Many-Particle Systems} (McGraw-Hill, San Francisco, 1971)

\bibitem{formally}
The ensemble average can only be performed \textit{formally} at this stage, because we do not know the quantity $\Lambda_{\bar{\sigma}}$, yet.
In fact, $\Lambda_{\bar{\sigma}}$ must be calculated \textit{self-consistently} from the formalism when the Green function is established. 
That is, $\Lambda_{\bar{\sigma}}$ both determines and is determined by the Green functions.
See Chap. 13 of Ref. \cite{fetter} for details.
This understanding is applicable to the quantity $\Delta$, which appears in the discussion of superconductivity.

\bibitem{tstarline}
$T^*$ line should be determined by either 1) all of the electrons in the system cannot `see' the energy gap ($n^*=n=1-\delta$) or 2) the energy gap disappears ($\Delta=0$, and therefore no electron can `see' the energy gap), whichever occurs first.
Our study shows that the condition 2), i.e., $\Delta=0$, occurs first.

\end{thebibliography}
\end{document}